# A search for a present-day candidate for the Comet P/Tunguska-1908

E.M.Drobyshevski[1], T.Yu.Galushina[2], M.E.Drobyshevski[1]

[1]) Ioffe Physical-Technical Institute, Russian Academy of Sciences, 194021 St-Petersburg, Russia;
   E-mail: emdrob@mail.ioffe.ru
[2]) Applied Mathematics and Mechanics Scientific Research Institute of Tomsk State University, Tomsk, Russia;
   E-mail: tanastra@mail.tomsknet.ru

The reason for the horizontal turn of the Tunguska-1908 bolide trajectory remains difficult to understand. It finds explanation, however, in the New Explosive Cosmogony of minor bodies as having been caused by an explosion of a part ($M$ up to $10^{12}$ g) of the comet nucleus whose ices contained products of its electrolysis, $2H_2+O_2$. In detonation, this part was repelled from the more massive unexploded nucleus remnant, changed the direction of its own motion by $\sim 10^o$ and imparted its kinetic energy, in expanding and slowing down, to the air in producing an effect of a high-altitude explosion. Because there are no traces of a fall of the more massive remnant, one comes to the conclusion that on passing through the Earth's atmosphere it again entered a heliocentric orbit (the hypothesis of V.Vernadskiy, 1932). A search for this comet, P/Tunguska-1908, among the 6077 known NEAs shows the 2005NB56 object to be the most appropriate candidate for a number of its parameters (a size is $\approx$ 170 m, $P$ = 2.106 y, $e$ = 0.473 and $i$ = 6.8$^o$). Back integration of its orbit made without allowing for non-gravitational effects suggests that it had passed the Earth on June 27, 1908 at a distance of 0.06945 AU. It is quite possible that a proper inclusion of even fairly weak non-gravitational forces would make its orbit fit in parameters that of the Tunguska bolide.

## 1. Introduction. The Tunguska-1908 phenomenon as a result of tangential passage of a cometary nucleus through the Earth's atmosphere

The Tunguska phenomenon (TP) of 1908 was triggered by a high-altitude ($\sim 5 \div 10$ km) atmospheric explosion of a body that was moving with $V \sim 20 \div 30$ km/s along a slightly sloping ($\delta \approx 0 \div 20^o$) trajectory with a kinetic energy $W \approx 10 \div 50$ Mt TNT (1 Mt TNT = $4.2 \times 10^{22}$ erg) (Krinov, 1949; Bronshten, 2000; Vasilyev, 2004; Longo, 2007). It is believed that this energy could transform to that of an explosion itself, i.e., the energy of the air overheated to $T > 10^4$ K, as it was slowing down the rapidly dispersing meteoroid.

Two major mechanisms of its breakup are mostly considered (see refs. in Drobyshevski, 2009), namely, (*i*) the so-called "explosion in flight", when the dynamic head of the air $\rho_a V^2/2$ exceeds the strength of the material (rock or ice) of the body, and (*ii*) chemical explosion of the cometary ices saturated with the solid solution of $2H_2+O_2$, the products of ice electrolysis (for details, see below) (the chemical explosion energy is less by far than $W$).

Studies of possible trajectories of the Tunguska meteoroid led Zotkin (1969), Kresák (1978), Asher and Steel (1998) to the conclusion that the TP could be caused by the fall of a fragment of the P/Encke nucleus ($P$ = 3.30 y, $a$ = 2.21 AU, $e$ = 0.85, $q$ = 0.34 AU, $i$ = 12$^o$) or of a meteoroid belonging to the β-Taurid stream ($P$ = 3.3 y, $e$ = 0.85, $q$ = 0.34 AU, $i$ = 6$^o$), with a maximum on June 30, visible from June 23 to July 7, which, in its turn, is believed to be genetically connected with P/Encke (Allen, 1973). Sekanina (1983, 1998) and Chyba *et al.* (1993) are more inclined to accept the asteroid nature of the TP. A statistical analysis of the problem led Farinella *et al.* (2001) to the conclusion that the TP was caused, with a probability of ≈83%, by an asteroid fall, and only with a probability of ≈17%, by the fall of the nucleus of an SP comet. In their recent work, Jopek *et al.* (2008) have also shown that the possible original trajectories of the Tunguska meteoroid resemble closer the asteroid than cometary orbits. Nevertheless, researchers studying the gas-dynamic aspects of the breakup and drag of the products of a meteoroid resulting from the above "explosion in flight" give preference to an icy body; indeed, ice is easier to break up because its strength is less than that of a rock, and, significantly, it evaporates without leaving behind, unlike an asteroid, rocky fragments - which is exactly the case of the TP.

The crucial aspect in unraveling the origin of the TP is the final turn of its trajectory. It is deduced both from the fact that (*i*) continuation of the original fireball trajectory with an azimuth $\varphi \approx 120°\div137°$ (reckoned clockwise from the direction to the north), as testified by numerous eyewitnesses, passed to the north of the final tree-fall, Kulik's epicenter (Bronshten, 2000; Vasilyev, 2004; Epiktetova, 2008) and that (*ii*) the azimuth of the symmetry axis $\varphi \approx 116°\div99°$ of this tree-fall, as also of the zone of radiation burn of the trees, deviates noticeably (at the very least, by $\Delta\varphi \approx 10°$) to the west of the direction of the original trajectory.

The concepts underlying the *New Explosive Cosmogony* (NEC) of minor bodies, which explains formation of SP cometary nuclei (and of a number of other bodies, of the type of MB asteroids, the Trojans, small satellites etc.) as due to extremely rare (7÷8 in 4.5 aeons) global explosions of thick (~800 km) electrolyzed icy envelopes of distant moonlike bodies, of the type of Ganymede or Titan, permitted Drobyshevski (2009) to show that all aspects of the TP, including the final turn of the trajectory, can be accounted for by detonation of the electrolysis products, $2H_2+O_2$ (present in the concentration of 15÷20 wt.%), dissolved in a part, or, better, in a layer up to $\sim 10^{12}$ g in mass (and $\sim 20 \times 200 \times 200$ m$^3$ in size) of a much more massive icy cometary nucleus. This yields ~200 m for the lower estimate of the size of the original nucleus. Judging from the visible size of the bolide (0.5÷2 km by Astapovich, 1934), the diameter of the meteoroid could be as large as ~500 m (which increases its mass to $\approx 10^{13} \div 10^{14}$ g).

The exploded layer was repelled with a velocity $V_r = 1.54 \div 1.63$ km/s from the much more massive nucleus, which accounts for the observed turn of the visible trajectory. The gaseous products of the layer's detonation expanding with $V_t \approx 2$ km/s were slowed down efficiently by the air, with the initial kinetic energy (up to $W \sim 50$ Mt TNT) and the momentum of the layer imparted to the air heated in the process to $T > 10^4$ K, and this is what created the phenomenon of a moving high-altitude TP explosion.

An essential point of the above scenario is the conclusion that the larger part of the cometary nucleus would be left intact in its tangential passage through the Earth's atmosphere, and a prediction that it would escape into space to enter again a heliocentric orbit. A similar, while hardly realizable possibility was mentioned as far back as 1932 by Vernadskiy, but Astapovich (1933) at about the same time replied that "There are at present no grounds to assume that a meteorite following a hyperbolic trajectory distorted by air resistance could escape beyond the confines of the Earth's atmosphere on passing through the perigee at Khatanga." As we see, now the NEC provides such grounds. Indeed, in as long as 100 years of the TP history only this concept has been able to offer, without invoking new hypotheses, a straightforward and physically transparent and noncontradictory interpretation of all known manifestations of the TP, including an explanation for the turn of the Tunguska fireball trajectory, a phenomenon which heretofore remained unexplained (and, therefore, commonly ignored or, conversely, used for concoction of various pseudo-scientific hypotheses).

The fall of the <u>main body</u> with a mass of $\approx 10^{13} \div 10^{14}$ g on the Earth with $W \sim 250 \div 3000$ Mt TNT would have produced a crater ~3.5÷8 km in diameter with the resultant climatic catastrophe of the type of the Younger Dryas cooling, which occurred ~13÷11.5 ky ago (see, e.g., Firestone *et al.*, 2007; Kennett *et al.*, 2008; and refs. in Drobyshevski, 2009). Fortunately, this fall did not occur, which suggests the only conclusion that the main body of the Tunguska bolide left the Earth on passing through the atmosphere at a distance of only ~10 km from its surface.

Because this icy body (let us call it P/Tunguska-1908 comet, or P/T subsequently), if it had not entered the gravitational sphere of other planets, can again strike the Earth, sooner or later, it appears only reasonable to make an attempt at locating it.

**2. From the trajectory of the Tunguska bolide to that of P/Tunguska-1908**

It would seem that the simplest approach to searching for the P/T would be, starting from the trajectory of the Tunguska fireball near the Earth's surface (within the confines of the Earth's atmosphere) and assuming only a small part of it to have exploded, to reconstruct its previous trajectory (see, e.g., Zotkin, 1969; Kresák, 1978; Asher and Steel, 1998; Jopek *et al.*, 2008; Zabotin and Medvedev, 2008) and project it, on a mathematically sound basis, up to the present time.

As initial conditions for the calculations, one could accept (with a number of reservations) the following data:

The SE → NW trajectory within the Earth's atmosphere had an azimuth $\varphi = 126°±12°$ (Epiktetova, 2008), the body passed through the point with $60°$ N, $106°$ E at an altitude $h_0 = 10$ km on June 30, 1908 at $t_0 = 0^h16'$ GMT. The body approached this point with a horizontal ($\delta = 0°$) velocity $V_0 \approx 20$ km/s (although the abovementioned authors used slightly different data sets, this produced only an insignificant effect on the conclusions of a fairly general nature concerning the original trajectory). At this point, a small part of the body exploded and imparted to it a transverse (lateral) velocity $V_1 \approx 130$ m/s in the ~NE direction.

One should take into account the air drag $F = c_x S \rho_a V^2/2$, where $\rho_a = \rho_0 \exp(-h/H_0)$, $\rho_0 = 0.0013$ g/cm$^3$, $H_0 = 7$ km, $c_x \approx 1$ is the hydrodynamic drag coefficient, $S = \pi \varnothing^2/4 = 7.1 \times 10^8$ cm$^2$, $M_{body} = 1.4 \times 10^{13}$ g.

An upper estimate of the air drag effect can be obtained by assuming that an icy body $\varnothing \approx 200$ m in size ($S = 3 \times 10^8$ cm$^2$) traversed with a velocity of 20 km/s without fragmentation at an altitude $h = 10$ km ($\rho_a = 0.4 \times 10^{-3}$ g/cm$^3$) a path $l = 1000$ km. The work done by the braking forces should have been $\approx 2.4 \times 10^{25}$ erg. A comparison of this value with the initial kinetic energy of $\sim 2 \times 10^{25}$ erg suggests that the gas dynamic braking over such a long path should be so efficient that the calculations must take into account the change in the body velocity caused by the braking.

Considered from this standpoint, the lateral change of the velocity of the main body by $V_1 \approx 130$ m/s caused by detonation of its small (~10%) side part may look not very significant, thus suggesting that some part of the transverse trajectory deviation could originate from a nonzero lift-drag ratio of an irregularly shaped body.

We see immediately that the original Tunguska-1908 meteoroid experienced so close encounter with the Earth and so complex gravitational manoeuvre accompanied by the partial explosion and by the atmospheric drag that it had to change crucially its initial orbit. Thus the starting parameters needed for calculation of the original present-day P/T orbit are known with errors so large as to be beyond the limits considered usually as acceptable, which makes their use in a search for this body in the present epoch hardly reasonable.

Our approach based on the NEC permits one *to invert the problem*, and in place of looking for the trajectory and other parameters of the original body, as this was done heretofore, we shall rather try to find now a body that, on sparing our cities a devastating Armageddon one hundred years ago, had sparked a brilliant warning fire show in the Siberian wilderness instead.

**3. Trajectory signature of the P/Tunguska-1908 comet**

The P/T1908 comet is an orbitally young object that has not yet had time enough to enter an orbit commensurate with that of the Earth. Assuming conventionally its orbital period $P \approx 3.5$ y ($a \approx 2.3$ AU) to be comparable to that of P/Encke (Zotkin, 1969; Kresák, 1978; Asher and Steel, 1998), then, in the absence of commensurability with the Earth, and disregarding perturbations caused by other planets, non-gravitational effects and the like, the probability for the residue of the Tunguska meteoroid to make a new passage through the Hill's gravitational sphere of the Earth ($\Delta$ = 0.01 AU) is about $3\times10^{-6}$ per year. Nevertheless, P/T approaches the Earth to a distance ≤1 AU once every ~25 y, when it may be discovered as an asteroid-like body, even if presently it is 'dormant' and does not reveal signs of cometary activity.

It is clear, however, that one should start with looking for the P/T among the already known near-Earth objects (NEOs) using the criteria summarized below.

It appears fairly obvious that

(1) A NEO-candidate for the P/T nucleus should cross the Earth's orbit sometime around June 30 (plus/minus a few days to account for the perturbations caused by other planets and probably non-gravitational forces initiated by persistent release of gas from areas of the nucleus surface which became exposed after tearing off the passive superficial shell and explosion of the nucleus part; for more details, see item 8 below).

(2) The distance to which the object approaches the Earth should not be large ($\Delta \leq 0.05$ AU); i.e., the search for P/T should be started among the Potentially Hazardous Asteroids (PHA by the Minor Planet Center terminology).

(3) Subsequent approaches (close to June 30) should occur with intervals from the TP date which are approximately multiples of the P/T orbital period.

(4) Back integration of the orbit of the P/T should demonstrate its close passage by the Earth (if not fall on it) on June 30, 1908.

(5) The orbit of the P/T, notwithstanding drastic perturbations which had occurred in passage through the Earth's atmosphere and explosion of a part of its nucleus, could nevertheless retain some orbital parameters of the original body (say, of a fragment of the P/Encke nucleus or of the member of the β-Taurid stream).

(6) The size of the P/T body is confined within ø ≈ 200÷500 m (the lower limit follows from the maximum estimate of the mass, ~$10^{12}$ g, of the exploded layer which changed its trajectory by ~5°÷10°; its mass should be much less than that of the original body, $M_i$, i.e., $M_i \approx 10^{13}$ g, whence for $\rho$ = 1 g/cm$^3$, ø ≈ 200 m; the upper limit is suggested by the estimate of Astapovich (1934), 0.5 km, of the minimal but nevertheless unusually large transverse size of the visible trace left by the original bolide).

(7) The P/T should most probably demonstrate some conventional cometary manifestations (a gaseous and a dust comas, flares etc.; see also item (1) above), which should, however, be about two orders of magnitude weaker than those observed with a typical SP comet, if not for any other reason than the nuclei differing in size by an order of magnitude (i.e., by two orders of magnitude in surface area).

(8) Accordingly, the non-gravitational forces which, generally speaking, may be considered being proportional to the surface area of the nucleus should impart to the small P/T nucleus an order-of-magnitude higher acceleration than to the nucleus of a conventional comet, because the size of the nuclei of the latter, as specified in the next paragraph, is measured in km.

For instance, the orbital period of P/Encke decreases after every revolution around the Sun, on the average, by 2.7 h (Whipple, 1950), which is equivalent to an averaged acceleration of ~$2\times10^{-6}$ cm/s$^2$. The radius of its nucleus is ~2.5÷4 km (Fernandez et al., 2000; Kelley et al., 2006; Boehnhardt et al., 2008); it is conceivable that for the same surface activity (and volume density) the P/T nucleus (with a radius of 100÷250 m) would feel an order-of-magnitude higher non-gravitational acceleration.

## 4. A Search for P/T among the Potentially Hazardous Asteroids

In accordance with item 2, Sec. 3, we started a search for candidates for the P/T with PHAs (see http://www.cfa.harvard.edu/iau/lists/PHACloseApp.html), i.e., among the bodies which are expected to approach the Earth to within $\Delta \leq 0.05$ AU in the time period from the present to the year 2178. Altogether, 27 objects satisfying the requirements of item 1 have been found, which make the closest approach around June 30 (±10 days). One more object could be added to this group, 153814 2001WN5, which comes close to the Earth, $\Delta = 0.0017$ AU, on June 26, 2028 (close competitors in this sense are Apophis with $\Delta = 0.00023$ AU and the quite recent object 2009DD45 with $\Delta = 0.00048$ AU) (Galushina, 2008), but was not listed among the PHAs for some reason.

Back integration performed with the use of NEODyS (http://newton.dm.unipi.it/neodys) codes revealed that none of these objects approached the Earth close enough on June, 1908. The closest objects were found to be 65909 1998FH12 with $\Delta = 0.179$ AU and the famous asteroid 25143 Itokawa (1998SF36) with $\Delta = 0.269$ AU. The object 153814 2001WN5 was at this time at $\Delta \approx 3.3$ AU, i.e., on the opposite side of its orbit. The integration with the MPC (http://www.cfa.harvard.edu) and the Tomsk (see Sec. 5 and Table 1 below) codes has yielded practically the same results (e.g., $\Delta = 0.1449$ AU for 65909 1998FH12 on 21 June, 1908, and $\Delta = 0.2422$ AU for 25143 Itokawa on 19 July, 1908, by the Tomsk calculations).

Absence in June, 1908, of close approaching objects among the known PHAs should not, however, be considered as a seriously discouraging one, because, as follows from the estimates given in the beginning of Sec. 3, the probability for the P/T to pass within $\Delta = 0.05$ AU from the Earth during 100 y is only ~0.01%.

**5. A search for the P/T among other NEAs**

The search was continued among all near-Earth asteroids (NEAs). This was effected by integration of the equations of motion of the 6077 NEAs back to January 1, 1908. The asteroid trajectories were integrated using the codes developed at the Applied Mathematics and Mechanics Scientific Research Institute of Tomsk State University (Galushina and Bykova, 2008; Bykova *et al.,* 2009). The starting data were taken from the catalogue of Bowell compiled up to January 12, 2009 (ftp:/ftp.lowell.edu/pub/elgb/astorb.dat). The integration was performed by the 19-order method of Everhart (Everhart, 1974). The model included the effects of all large planets, Pluto and the Moon. The coordinates of the perturbing bodies were calculated on the basis of the DE405 ephemerises.

Table 1. Asteroids which approached the Earth close to June 30, 1908. $\Delta$ - the approach distance; $P$ - period of revolution around the Sun; $a, e, i$ - major semiaxis, eccentricity and orbital inclination; $q, Q$ - perihelion and aphelion distances; $H$ - stellar magnitude; $d$ - size of the object (for albedo 0.04; except Itokawa).

| Name | 25143 Itokawa | 65909 1998 FH12 | 2005 MB | 2005 NB56 |
|---|---|---|---|---|
| $\Delta$, AU | 0.2728 (30.06.1908) | 0.1746 (30.06.1908) | 0.0971 (26.06.1908) | 0.06945 (27.06.1908) |
| $P$, days | 556.4798 | 416.4360 | 357.2113 | 769.2369 |
| $a$, AU | 1.3240 | 1.0914 | 0.9853 | 1.6430 |
| $e$ | 0.2800 | 0.5397 | 0.7928 | 0.4728 |
| $i$, deg | 1.6219 | 3.5585 | 41.4163 | 6.7633 |
| $q$, AU | 0.9533 | 0.5024 | 0.2042 | 0.8661 |
| $Q$, AU | 1.6948 | 1.6804 | 1.7663 | 2.4199 |
| H | 19.2 | 19.2 | 17.07 | 22.94 |
| $d$, m | 520×270×230 | 970 | 2600 | 170 |

Our study revealed that two objects, 2005MB ($\Delta = 0.0971$ AU, Aton) and 2005NB56 ($\Delta = 0.0695$ AU, Apollo), approached within this selected model fairly close the Earth at the end of June 1908. We subjected the motion of these asteroids and of the two PHA objects mentioned in the preceding Sec. 4 to a more comprehensive study. The force model was complemented by the Earth's oblatness and three asteroids (Ceres, Pallas, Vesta). Table 1 lists relevant data on all the four objects.

We are planning to continue our studies of the motion of these objects. To more accurately compute the asteroid positions at the end of June 1908, we will have to study the probability motions domain and try to take into account the non-gravitational forces.

**6. Main conclusions. Is 2005NB56 the P/Tunguska-1908?**

As seen from Table 1, the most appropriate candidate for P/T among the known NEOs is the 2005NB56 asteroid (interestingly, only 5 years ago, when it had not yet been discovered, this study could not have brought such a strong statement).

This statement is corroborated not only by (1) the closeness of the date (June 27, 1908) to the time of the TP and (2) the smallest distance to the Earth among the four objects found ($\Delta = 0.06945$ AU) at the time, but (3) by its orbital parameters being closest to those of P/Encke and of the β-Taurid stream and, finally and most significantly, (4) by the size of the object (≈170 m) practically coinciding with the lower estimate derived from the angle of turn of the Tunguska fireball trajectory (Drobyshevski 2009), while the dimensions of the other three objects exceed by far the value derived from the TP observations.

There is a problem in that back integration of the orbit of the 2005NB56 object does not terminate at the exact location of the TP event. The matter is that these calculations disregard the forces of other than gravitational origin which should inevitably manifest themselves in our case, where we deal with a remnant of a cometary nucleus. As pointed out in Sec. 3, one can, in principle, expect non-gravitational forces to impart the P/T nucleus an acceleration exceeding noticeably the non-gravitational acceleration of the P/Encke nucleus. One can readily verify that in order to change the position of a body by $\Delta = 0.06945$ AU $\approx 10^{12}$ cm in 100 y, its acceleration with respect to the orbital velocity vector should be maintained constant at a level of $\sim 2\times 10^{-7}$ cm/s$^2$, which is about an order of magnitude smaller than that experienced, for example, by the nucleus of P/Encke. It is worth noting here that non-gravitational parameters are empirical quantities. They undergo sometimes fast and unpredictable variations (see, e.g., Pittich *et al.,* 2004, and refs. therein), because they originate from reactive action of nonstationary gas/dust jets ejected from the small-area sources on the nucleus. On the other hand, it is quite possible that such jets from the P/T nucleus left unexploded (i.e., strongly depleted in the volatiles and $2H_2+O_2$) did not last for long and decayed in a few years or tens of years, thus making, nevertheless, impossible exact calculation of the position of the comet in the past. One cannot exclude the possibility that careful observations of the 2005NB56 object when it will approach the Earth closer (see Table 2), will show it to reveal certain cometary manifestations (regrettably, in June 2009, its $\Delta \approx 1.5$ AU).

Table 2. Close approaches of 2005NB56 in future

| Date | $\Delta$, AU |
|---|---|
| 2045.07.11 | 0.04249 |
| 2064.06.22 | 0.09505 |
| 2123.07.06 | 0.03090 |
| 2163.06.25 | 0.09344 |

It thus appears that the argument about the origin of the object responsible for the TP lacks to some extent the essence (see Introduction). Judging from the observation that after its passage by the Earth it continued its motion in an asteroid orbit, this was indeed an object belonging to the asteroid population (i.e., all these authors, viz., Sekanina, 1983, 1998;

Farinella *et al.,* 2001; Jopek *et al.,* 2008, are right). But P/Encke also moves in an asteroid-like orbit. And the problem of how to transfer P/Encke from a well-behaving Jupiter's family into an asteroid orbit is old and far from straightforward (e.g., Asher and Steel, 1998; Pittich *et al.,* 2004; Jopek *et al.,* 2008).

Strange though this might seem, the NEC introduces some order here too by removing fictitious contradictions borne by obsolete paradigms. Indeed, by the NEC the MB asteroids originated from an explosion of a Pluto-like icy planet (Drobyshevski, 1980, 1986) (followed by a collision of the retained rocky planetary core with the planet's satellite lost due to the explosion (Drobyshevski, 1997), an event that can be correlated with origin of the highly inclined Pallas orbit and that one can identify with the indications of the second shock metamorphism observed in minerals of the well-known meteorites by the professional geologist Rezanov (2004)). It is these events (~3.9 aeons ago they apparently brought about the Imbrium bombardment of the Moon) that produced the rocky and iron asteroids, as well as icy fragments, which are nuclei of originally active comets. A non-central explosion of the dwarf planet is capable of accounting for the observed distribution of asteroids of a variety of types in their distance from the Sun (Drobyshevski *et al.,* 1994). The icy fragments, their surface layers outgassed long ago, became low-albedo (~0.05) C-type asteroids, but many of them did certainly retain in their bulk ice with dissolved $2H_2+O_2$ to make them 'dormant' comets. Collisions with other asteroid bodies, which are not so rare in the Main Belt, are capable of activating such a sleeping comet; this is possibly the case with P/Encke which, as we see, may possibly never have been a member of the Jupiter family. Belonging to this class was the Tunguska meteoroid, which accounts for all the TP manifestations and suggests certain inferences concerning the origin of the 2005NB56 NEO and of similar objects.

**References**

Allen, C.W., 1973. *Astrophysical Quantities,* 3rd Ed., The Athlone Press, London.

Asher, D.J., Steel, D.I., 1998. On the possible relation between the Tunguska bolide and comet Encke. *Planet. Space Sci.* **46**(2/3), 205-211.

Astapovich, I.S., 1933. New data about the fall of the great meteorite on June 30, 1908, in Central Siberia. *Astron. Zh.* **10**(4), 465-486.

Astapovich, I.S., 1934. Air waves caused by the fall of the meteorite on 30th June, 1908, in Central Siberia. *Q. J. Roy. Meteorol. Soc*. **60**, No257, 498-504.

Boehnhardt, H., Tozzi, G.P., Bagnulo, S., Muinonen, K., Nathues, A., Kolokolova, L., 2008. Photometry and polarimetry of the nucleus of comet 2P/Encke. *Astron. Astrophys.* **489**, 1337-1343; arXiv:0809.1959.

Bronshten, V.A., 2000. *Tungusskiy Meteorit: Istorija Issledovanija(Tunguska Meteorite: History of Studies)*. Publ. Seljanov A.D., Moscow.

Bykova, L.E., Galushina, T.Yu., Baturin, A.P., 2009. The algorithms and programs for investigations of near-Earth asteroids. *Astron. Astrophys. Trans*. (in press).

Chyba, C.F., Thomas, P.J., Zahnle, K.J., 1993. The 1908 Tunguska explosion: Atmospheric disruption of a stony asteroid. *Nature* **361**, 40-44.

Drobyshevski, E.M., 1980. Electrolysis in space and fate of Phaethon. *Moon Planets* **23**, 339-344.

Drobyshevski, E.M., 1986. The structure of Phaethon and detonation of its icy envelope. *Earth Moon Planets* **34**, 213-222.

Drobyshevski, E.M., 1997. The origin of the asteroid main belt: Synthesis of mutually exclusive paradigms. *Astron. Astrophys. Trans*. **12**, 327-331.

Drobyshevski, E.M., 2009. Tunguska-1908 and similar events in light of the New Explosive Cosmogony of minor bodies. arXiv:0903.3309.

Drobyshevski, E.M., Simonenko, V.A., Demyanovski, S.V., Bragin, A.A., Kovalenko, G.V., Shnitko, A.S., Suchkov, V.A., Vronski, A.V., 1994. New approach to the explosive origin of the asteroid belt. In: Kozai, Y., Binzel, R.P., Hirayama, T. (Eds.), *Seventy-Five Years of Hyrayama Asteroid Families*, ASP Conference Series, **63**, pp.109-115.

Epiktetova, L.E., 2008. Trajectory of Tunguska Cosmic Body on the basis of witness testimonies. In: Intnl. Conf. "*100 years since Tunguska Phenomenon: Past, Present and Future*" (Abstracts), Moscow; pp.77-79.

Everhart E., 1974. On Efficient Integrator of Very High Order and Accuracy with Appendix Listing of RADAU. Denver, Univ. of Denver, p.20.

Farinella, P., Foschini, L., Froeschlé, Ch., Gonczi, R., Jopek, T.J., Longo, G., Michel, P., 2001. Possible asteroidal origin of the Tunguska cosmic body. *Astron. Astrophys.* **377**, 1081-1097.

Fernandez, Y.R., Lisse, C.M., Ulrich Kaufl, H., Peschke, S.B., Weaver, H.A., A'Hearn, M.F., Lamy, P.P., Livengood, T.A., Kostiuk, T., 2000. Physical properties of the nucleus of comet 2P/Encke. *Icarus* **147**(1), 145-160.

Firestone, R.B., West, A., Kennett, J.P., Becker, L., Revay, Z.S., Schultz, P.H., Belgya, T., Kennett, D.J., Erlandson, J.V., Dickenson, O.J., Goodyear, A.C., Harris, R.S., Howard, G.A., Kloosterman, J.B., Lechler, P., Mayewski, P.A., Montgomery, J., Poreda, R., Darrah, T., Que Hee, S.S., Smith, A.R., Stich, A., Topping, W., Wittke, J.H., Wolbach, W.S., 2007. Evidence for an extraterrestrial impact 12,900 years ago that contributed to the megafaunal extinctions and the Younger Dryas cooling. *Proc. Natl. Acad. Sci. U.S.A*. **104**, 16016-16021.

Galushina, T.Yu., 2008. The investigation of motion of the asteroids coming upon gravitation sphere of the Earth in next 120 years. In: Intnl. Conf. "*100 years since Tunguska Phenomenon: Past, Present and Future*" (Abstracts), Moscow; pp.109-110.

Galushina T., Bykova L., 2008. Applied program system for study of motion and orbital evolution of asteroids. In: Orlov, V.V., Rubinov, A.V. (Eds.), "*Resonances, Stabilization, and Stable Chaos in Hierarchical Triple Systems*". St. Petersburg University, pp.5-11.

Jopek, T.J., Froeschlé, C., Gonczi, R., Dybczyński, P.A., 2008. Searching for the parent of the Tunguska cosmic body. *Earth Moon Planets* **102**(1-4), 53-58.

Kelley, M.S., Woodward, C.E., Harker, D.E., Wooden, D.H., Gehrz, R.D., Campins, H., Hanner, M.S., Lederer, S.M., Osip, D.J., Pittichova, J., Polomski, E., 2006. A *Spitzer* Study of Comets 2P/Encke, 67P/Churyumov-Gerasimenko, and C/2001 HT50 (LINEAR-NEAT). *Astrophys. J.* **651**, 1256-1271; arXiv:astro-ph/0607416.

Kennett, D.J., Kennett, J.P., West,G.J., Erlandson, J.M., Johnson, J.R., Hendy, I.L., West, A., Culleton, B.J., Jones, T.L., Stafford, Jr., T.W., 2008. Wildfire and abrupt ecosystem disruption on California's Northern Channel Islands at the Ållerød-Younger Dryas boundary (13.0-12.9 ka). *Quat. Sci. Rews.* **27**, 2528-2543.

Kresák, L'., 1978. The Tunguska object: a fragment of comet Encke? *Bull. Astron. Inst. Czechosl*. **29**, 129-134.

Krinov, E.L., 1949. *Tungusskiy Meteorit*, Publ. USSR Acad. Scis., Moscow.

Longo, G., 2007. The Tunguska event. In: Bobrowsky, P.T., Rickman, H. (Eds.), *Comet/Asteroid Impacts and Human Society. An Interdisciplinary Approach*. Springer Verlag, Heidelberg - New York; Chap. 18, pp.303-330.

Pittich, E.V., D'Abramo, G., Valsecchi, G.B., 2004. From Jupiter-family to Encke-like orbits. The rôle of non-gravitational forces and resonances. *Astron. Astrophys.* **422**, 369-375.

Rezanov, I.A., 2004. *Istoriya Vzorvavsheysya Planety (History of Exploded Planet)*. Rus. Acad. Scis., Moscow.

Sekanina, Z., 1983. The Tunguska event: No cometary signature in evidence. *Astron. J.* **88**, 1382-1414.

Sekanina, Z., 1998. Evidence for asteroidal origin of the Tunguska object. *Planet. Space Sci.* **46**(2/3), 191-204.

Vasilyev, N.V., 2004. *Tunguska Meteorite: Space Phenomenon of 1908 Summer*. Publ. NP ID "Russkaja Panorama", Moscow.

Vernadskiy, V.I., 1932. On study of space dust. *Mirovedenie* **21**(5), 32-41.

Whipple, F.L., 1950. A comet model. I. The acceleration of comet Encke. *Astrophys. J.* **111**, 375-394.

Zabotin, A.S., Medvedev, Yu.D., 2008. On possible heliocentric orbit of the Tungus meteorite before entrance to the Earth's atmosphere. In: Rykhlova, L.V., Tarady, V.K. (Eds.), *Near-Earth Astronomy 2007*, Nalchik; pp.149-154.

Zotkin, I.T., 1969. Anomalous twilight related with the Tunguska meteorite. *Meteoritika* **29**, 170-176.